\documentclass[usenatbib]{mnras}
\usepackage[T1]{fontenc}
\usepackage[latin9]{inputenc}
\usepackage{bm}
\usepackage{amsmath}
\usepackage{graphicx}
\begin{document}
\title{Shape of solar cycles and mid-term solar activity oscillations}
\author[D.D.~Sokoloff,A.S. Shibalova, V.N. Obridko, V.V.~Pipin]
{D.~Sokoloff$^{1,2,3}$\thanks{email: sokoloff.dd@gmail.com},
  A.S. Shibalova$^{2}$,\thanks{email:as.shibalova@physics.msu.ru},
  V.N. Obridko$^{2}$, \thanks{email:obridko@izmiran.ru},
V.~Pipin$^{4}$\thanks{email: pip@iszf.irk.ru}\\
$^{1}${Department of Physics, Moscow State University,
  Moscow,119992, Russia}\\
$^{2}${IZMIRAN, 4 Kaluzhskoe  Shosse, Troitsk, Moscow, 142190}\\
$^{3}$Moscow Center of Fundamental and Applied Mathematics, Moscow,
119991, Russia\\
$^4${Institute of Solar-Terrestrial Physics, Russian Academy of
Sciences, Irkutsk, 664033, Russia} }
\maketitle

\begin{abstract}
The evolution of the solar activity comprises, apart from the well-known
11-year cycle, various temporal scales ranging from months up to the
 secondary cycles known as mid-term oscillations. Its nature deserves a physical
explanation. In this work, we have considered the 5-to-6 year oscillations
as derived both from sunspot and from solar magnetic
dipole time series. Using solar dynamo model, we have deduced that these
variations may be a manifestation of the dynamo nonlinearities and
non-harmonic shape of the solar activity cycles. We have concluded
that the observed mid-term oscillations are related to the nonlinear
saturation of the dynamo processes in the solar interior. 
\end{abstract}

\section{Introduction}

The most pronounced feature in the time evolution of the solar activity
is the well-known 11-year sunspot cycle with the 22-year oscillations of the solar
magnetic field polarity underlying it. The evolution comprises also  various shorter temporal scales known as
mid-term oscillations (MTO) that range from months up to several years. 
The best known MTO are the so-called
quasi-biennial oscillations, QBO (see, for example, the comprehensive review by
\cite{Baz2014} and references therein). The periodicities in this
range are sometimes referred to as intermediate- or mid-term quasi-periodicities (e.g., \cite{Letal03}).
 Another mid-term periodicity that was reported is the six-year oscillations \cite{Prab02}. 
For this phenomenon, we propose the term Quasi-Sexennial Oscillations (QSO). These two components
of MTO (QBO and QSO) have different properties, and perhaps, different origins.

In this paper, we concentrate mainly on the analysis of QSO. 
Experimental data on QSO are rather scarce. This is because 
the periods of 4-6 years are difficult to isolate from the full spectrum
in the vicinity of a powerful 11-year spectral peak. 
One way to improve the reliability of the detection of QSO parameters is to use a time
series that has at least 100 years of data. 
The sunspot number record has one of the longest available time series of the solar activity tracers. 
The results of \cite{Frick2020} show that mid-term oscillations (including QBO and QSO) in the sunspot
numbers can be considered elements of the continuous spectrum typical
of various turbulent or convective systems. Nevertheless, some features and characteristics
of this spectrum can be originated in a deterministic way, e.g., due to nonlinearities in the large-scale dynamos.

According to \cite{Parker1955}, the solar dynamo involves two major components of the magnetic activity. 
The large-scale toroidal field produces bipolar regions, which emerge in the form of sunspots, 
and the large-scale poloidal field forms the shape of the solar corona. Correspondingly, we
consider  QSO both in sunspot tracers and in the tracers of the global magnetic field observed on the solar surface. 
Based on a joint analysis of the sunspot number, properties of the solar-wind plasma, interplanetary magnetic field, and 
geomagnetic activity index Ap, \cite{Prab02} found
variations with periods of about 5.5 years. 
\cite{Deng14} identified variations
with a period of 4-5 years while analyzing the characteristics of polar plages. 
Oscillations in the QSO range were later re-discovered by \cite{Deng20}
from data on the solar corona rotation over 80 years. 
Early on, \cite{Riv92} found
variations with a period of about 6 years by analyzing the spectrum
of the magnetic field of the Sun as a star based on the data of the
John Wilcox Observatory in Stanford. \cite{Liv06} studied separately
the cyclic variations of the axial and equatorial dipoles of the Sun.
They found out that 5-year variations were observed only in the axial dipole record.

 In search for an explanation of the MTO phenomenon, we
 recall that the continuous turbulent spectrum in the Kolmogorov turbulence
is connected with the presence of the nonlinear transport term $\bm{v}\cdot\nabla\bm{v}$
in the Navier-Stokes equation. Because of this term, a large vortex
of scale $l$ produces two vortexes of a smaller scale $l/2$, while
the vortex wave vector $k$ produces the wave vector $2k$. This is
a standard explanation for the nature of the turbulent cascade. This explanation
is, however, not immediately applicable to the case under discussion.
Moreover, the proper oscillation scale for the magnetic field is 22
years, while 11 years is only the oscillation scale for the magnetic energy.

On the other hand, quadratic terms are present in such a problem.
For example, the magnetic force acting on the flows and affecting the dynamo efficiency 
is quadratic with respect to the magnetic field. Also, the five years variation is, indeed, 
about half the duration of the 11-yr solar cycle. 
It seems plausible that something similar to a turbulent cascade may take place in the 
solar activity engine. Of course, this analogy is a very rough hint but most certainly worthy
of a proper investigation here. This is the aim of our paper.

\section{Observational basis}

In the present work, we are using the sunspot numbers and
 the parameters of the dipole term in the multipole expansion of the 
surface large-scale magnetic field as observational characteristics 
of the solar activity variations.

The raw monthly mean sunspot numbers SSN were borrowed 
from WDC-SILSO summary data, Royal Observatory of Belgium, Brussels
 (http://sidc.oma.be/silso/datafiles, version 2)

Using the Wilcox Solar Observatory (WSO) synoptic charts
of the radial component of the solar magnetic field \cite{SW77}, 
we calculated the magnetic field under the potential approximation by the well-known method 
described in \cite{HS86} and \cite{H91}. The method was applied 
in its classical version, without assuming radial field in the photosphere.
 WSO measurements of the magnetic-field longitudinal component were used as 
the source data to plot the synoptic charts for each Carrington rotation.
 The WSO data used in this study cover a time interval of 43 years 
from the beginning of Carrington rotation (CR) 1642 (27 May 1976) to the end of CR 2210 (November 2018).

The magnetic field parameters were calculated by solving the boundary-value problem
 for the line-of-sight field component measured in the photosphere and a strictly radial field
 at the source surface. The latter is assumed to be located at a distance of 2.5 radii from the center of the Sun.

The results are expressed in terms of the expansion in the Legendre polynomials.
 The terms $g_{lm}$ and $h_{lm}$ are the coefficients of the spherical harmonic analysis obtained 
by comparison with observations at the photospheric level. 
It is important to note that the coefficients were calculated under the assumption that the field is 
potential throughout the photosphere up to the source surface, including the boundaries. 
These coefficients determine the contribution of various multipoles and their direction.
 The dipole magnetic moment $M_{{\rm dip}}$, the axial dipole moment $M_{{\rm ax}}$, and 
the equatorial dipole moment $M_{{\rm eq}}$ are determined by the following equations:
\begin{eqnarray}
M_{{\rm dip}}=\sqrt{g_{10}^{2}+g_{11}^{2}+h_{11}^{2}},\\
M_{{\rm ax}}=|g_{10}|,\quad\quad M_{{\rm eq}}=\sqrt{g_{11}^{2}+h_{11}^{2}}.
\end{eqnarray}

As a first step in this paper, we will discuss variations in the solar dipole, 
since it is most simply related to the main mechanism of generation of the magnetic field. 
It is easy to model and its oscillations have the simplest form. 
Despite its simplicity, the dipole demonstrates the main features characteristic 
of the general description of solar activity: 
 the solar dipole magnetic field displays both axisymmetric and non-axisymmetric components, 
whose variations with latitude and phase of the cycle obey the basic laws of solar activity as a whole.

\section{The search for MTO using wavelet analysis}

The Fourier analysis is a traditional way to search for periodicity in 
stationary, strictly periodic or localized signals. 
The wavelet analysis has been applied to take into account quasi-periodic processes
 and time-localized periodicities (see e.g., \cite{Frick2020} in the context of solar activity problems). 
This method compares the signal with localized waves unlike the harmonic decomposition with 
an infinite sinusoidal signal. The wavelet existence time depends on two parameters - 
the wave position $t$ and the characteristic scale $a$.

The general wavelet transform of function $f(t)$ is defined as 
\begin{equation}
W(a,t)=C_{\psi}^{-1/2}a^{-1/2}\int\limits _{-\infty}^{\infty}\psi^{*}\left(\frac{t'-t}{a}\right)f(t')dt',\label{eq1}
\end{equation}
where $\psi(t)$ means the basic wavelet function, {*} denotes a complex
conjugation, and coefficient $C_{\psi}$ is determined through a Fourier
transform of the basic wavelet $\hat{\psi}(\omega)$: 
\begin{equation}
C_{\psi}=\int\limits _{-\infty}^{\infty}|\omega|^{-1}|\hat{\psi}(\omega)|^{2}d\omega.\label{eq2}
\end{equation}
The most widely used wavelets are derivatives of the Gaussian function.
In our paper, we consider one of them --- a complex valued Morlet
wavelet 
\begin{equation}
\psi(t)=e^{-t^{2}/\alpha^{2}}e^{i\omega t},\label{eq3}
\end{equation}
with the parameters $\alpha^{2}=2$ and $\omega=2\pi$.

The global wavelet spectrum 
\begin{equation}
S(a)=\int|W(a,t)|^{2}dt\label{eq4}
\end{equation}
shows the scale distribution of energy.

The reliability of the wavelet analysis depends on the length of the data record. 
It is usually sufficient if the realization is at least three times longer than 
the expected oscillation period. In our case, the length of the magnetic-field database (43 years) ensures 
the reliability of the detected periods no longer than 11-15 years. 
Therefore, the results obtained here for Mid-Term Oscillations (MTO) are quite reliable.
 Furthermore, the sunspot data we are using cover a period of more than 200 years. 
Therefore, the given sets of data are sufficient to reliably characterize the MTO parameters.

\section{Dynamo model}

In this paper, we are using the dynamo model proposed by \cite{Pipin2018e}.
 It includes the minimal set of the dynamo equations to model the evolution of the non-axisymmetric magnetic field.
 Similarly to \cite{Moss2008}, we neglect the radial dependence of the magnetic field and
 assume that the radial gradient of the angular
velocity is larger than the latitudinal gradient. The evolution of the
large-scale magnetic field induction vector $\left\langle \mathbf{B}\right\rangle $
in a perfectly conductive medium is governed by the mean-field equation
\begin{equation}
\partial_{t}\left\langle \mathbf{B}\right\rangle =\boldsymbol{\nabla}\times\left(\boldsymbol{\mathcal{E}}+\left\langle \mathbf{U}\right\rangle \times\left\langle \mathbf{B}\right\rangle \right),\label{eq:mfe-1}
\end{equation}
where, $\boldsymbol{\mathcal{E}}=\left\langle \mathbf{u\times b}\right\rangle $
is the mean electromotive force with $\mathbf{u}$ and $\mathbf{b}$
standing for the turbulent velocity and magnetic field, respectively.
It is convenient to represent the vector $\left\langle \mathbf{B}\right\rangle $
in terms of the axisymmetric and non-axisymmetric components as follows:
\begin{eqnarray}
\left\langle \mathbf{B}\right\rangle  & = & \overline{\mathbf{B}}+\tilde{\mathbf{B}}\label{eq:b0}\\
\mathbf{\overline{B}} & = & \hat{\boldsymbol{\phi}}B+\nabla\times\left(A\hat{\boldsymbol{\phi}}\right)\label{eq:b1}\\
\tilde{\mathbf{B}} & = & \boldsymbol{\nabla}\times\left(\mathbf{r}T\right)+\boldsymbol{\nabla}\times\boldsymbol{\nabla}\times\left(\mathbf{r}S\right),\label{eq:b2}
\end{eqnarray}
where $\overline{\mathbf{B}}$ and $\tilde{\mathbf{B}}$ are the axisymmetric
and non-axisymmetric components; ${A}$, ${B}$, ${T}$, and ${S}$
are the scalar functions representing the field components; $\hat{\boldsymbol{\phi}}$
is the azimuthal unit vector, $\mathbf{r}$ is the radius vector;
$r$ is the radial distance, and $\theta$ is the polar angle. Hereafter,
the overbar denotes the axisymmetric magnetic field, and the tilde
denotes non-axisymmetric properties. 

Following the ideas outlined above, we now
consider a reduced dynamo model where we neglect the radial dependence
of the magnetic field. In this case, the induction vector of the large-scale
magnetic field is represented in terms of the scalar functions as
follows: 
\begin{eqnarray}
\left\langle \mathbf{B}\right\rangle  & = & -\frac{\mathbf{r}}{R^{2}}\frac{\partial\sin\theta A}{\partial\mu}-\frac{\hat{\theta}}{R}A+\hat{\boldsymbol{\phi}}B\label{eq:3d}\\
 & - & \frac{\mathbf{r}}{R^{2}}\Delta_{\Omega}S+\frac{\hat{\theta}}{\sin\theta}\frac{\partial T}{\partial\phi}+\hat{\phi}\sin\theta\frac{\partial T}{\partial\mu},\nonumber 
\end{eqnarray}
where $R$ represents the radius of the spherical surface inside a
star where the hydromagnetic dynamo operates. The above equation defines
the 3d divergence-free B-field on the sphere. The model employs the
following expression for $\boldsymbol{\mathcal{E}}$:

\begin{eqnarray}
\boldsymbol{\mathcal{E}} & = & \alpha\left\langle \mathbf{B}\right\rangle -\eta_{T}\boldsymbol{\nabla}\times\mathbf{\left\langle B\right\rangle }+V_{\beta}\hat{\mathbf{r}}\times\mathbf{B}.\label{eq:simpE}
\end{eqnarray}

Applying these simplifications to Eq (\ref{eq:mfe-1}) and Eqs (\ref{eq:b0}-\ref{eq:b2}),
we obtain the following set of dynamo equations in terms of the scalar
functions, $A,B,S$, and $T$: 
\begin{eqnarray}
\partial_{t}B & =- & \sin\theta\frac{\partial\Omega}{\partial r}\frac{\partial\left(\sin\theta A\right)}{\partial\mu}+\eta_{T}\frac{\sin^{2}\theta}{R^{2}}\frac{\partial^{2}\left(\sin\theta B\right)}{\partial\mu^{2}}\nonumber \\
 &  & +\frac{\sin\theta}{R}\frac{\partial}{\partial\mu}\alpha_{0}\mu\left\langle B_{r}\right\rangle +\frac{\alpha_{0}\mu}{R}\left\langle B_{\theta}\right\rangle \label{eq:bt}\\
 &  & -\frac{1}{R}V_{\beta}\left\langle B_{\phi}\right\rangle -\frac{B}{\tau}\nonumber 
\end{eqnarray}
\begin{equation}
\partial_{t}A=\alpha_{0}\mu\left\langle B_{\phi}\right\rangle +\eta_{T}\frac{\sin^{2}\theta}{R^{2}}\frac{\partial^{2}\left(\sin\theta A\right)}{\partial\mu^{2}}-\frac{V_{\beta}}{R}A-\frac{A}{\tau},\label{eq:at}
\end{equation}
To obtain the evolution equations for potential $S$ and $T$, we follow
the procedure described in detail in \cite{Krause1980}.
We get, 
\begin{eqnarray}
\partial_{t}\Delta_{\Omega}T & = & -\Delta_{\Omega}\delta\Omega\frac{\partial T}{\partial\phi}+\frac{\eta_{T}}{R^{2}}\Delta_{\Omega}^{2}T\label{eq:Tt}\\
 & - & \frac{1}{R}\frac{\partial\Omega}{\partial r}\sin^{2}\theta\frac{\partial\Delta_{\Omega}S}{\partial\mu}-\frac{1}{R}\frac{\partial}{\partial\phi}\left[\frac{\alpha_{0}}{\sin\theta}\mu\left\langle B_{\phi}\right\rangle \right]\nonumber \\
 & + & \Delta_{\Omega}\frac{\alpha_{0}\mu}{R}\left(\left\langle B_{r}\right\rangle \sin^{2}\theta+\mu\sin\theta\left\langle B_{\theta}\right\rangle \right)\nonumber \\
 & + & \frac{1}{R}\frac{\partial}{\partial\mu}\alpha_{0}\mu\sin\theta\left\{ \mu\sin\theta\left\langle B_{r}\right\rangle +\mu^{2}\left\langle B_{\theta}\right\rangle \right\} \nonumber \\
 & - & \frac{1}{R\sin\theta}\frac{\partial}{\partial\phi}\left\langle B_{\theta}\right\rangle V_{\beta}-\frac{\partial}{\partial\mu}\left(\sin\theta\left\langle B_{\phi}\right\rangle V_{\beta}\right),\nonumber 
\end{eqnarray}
\begin{eqnarray}
\partial_{t}\Delta_{\Omega}S & = & -\left(\delta\Omega\Delta_{\Omega}\frac{\partial}{\partial\phi}S\right)+\frac{\eta_{T}}{R^{2}}\Delta_{\Omega}^{2}S\label{eq:St}\\
 &  & +\frac{\partial}{\partial\mu}\alpha_{0}\mu\sin\theta\left\langle B_{\phi}\right\rangle \nonumber \\
 & + & \frac{\partial}{\partial\phi}\left\{ \frac{\alpha_{0}\mu}{\sin\theta}\left(\left\langle B_{\theta}\right\rangle +\sin\theta\left(\mathbf{e}\cdot\left\langle \mathbf{B}\right\rangle \right)\right)\right\} \nonumber \\
 & - & \frac{1}{\sin\theta}\frac{\partial}{\partial\phi}\left(\left\langle B_{\phi}\right\rangle V_{\beta}\right)+\frac{\partial}{\partial\mu}\left(\sin\theta\left\langle B_{\theta}\right\rangle V_{\beta}\right),\nonumber 
\end{eqnarray}
where ${\displaystyle \Delta_{\Omega}=\frac{\partial}{\partial\mu}\sin^{2}\theta\frac{\partial}{\partial\mu}+\frac{1}{\sin^{2}\theta}\frac{\partial^{2}}{\partial\phi^{2}}}$
and $\mu=\cos\theta$. To simulate the stretching of non-axisymmetric
magnetic field by the surface differential rotation, we consider the
latitudinal dependence of the angular velocity $\delta\Omega=-0.25\sin^{2}\theta\Omega$
in Eqs (\ref{eq:Tt}) and (\ref{eq:St}), which are written in the
coordinate system rotating with angular velocity $\Omega$. {
}The $\tau$-terms in Eqs(\ref{eq:bt},\ref{eq:at}) were suggested
by \cite{Moss2008} to account for turbulent diffusion in the radial
direction. Similarly to the cited paper,
 we put { ${\displaystyle \tau=3\frac{R^{2}}{\eta_{T}}}$}.
The magnetic buoyancy is the source of the non-axisymmetric magnetic field in the model. 
We assume that the magnetic buoyancy affects relatively small-scale parts of the axisymmetric magnetic field, 
perhaps, because of some kind of nonlinear instability, and contributes to the induction of 
the non-axisymmetric magnetic field component. 
In the mean-field models, the magnetic buoyancy acts to produce a nonlinear turbulent pumping effect.
 Following \cite{Kitchatinov1993}, we have:
\begin{eqnarray}
V_{\beta}=\begin{cases}
\dfrac{\alpha_{MLT}u'}{\gamma}\beta^{2}K\left(\beta\right)\left[1+\xi_{\beta}\left(\phi\right)\right], & \text{if}\;\beta\ge\beta_{cr},\\
0, & \text{if}\;\beta<\beta_{cr}
\end{cases}\label{eq:buoy}
\end{eqnarray}
where $\beta=\left|\left\langle \mathbf{B}\right\rangle \right|/\mathrm{B_{eq}}$,
$\mathrm{B_{eq}}=\sqrt{4\pi\overline{\rho}u'^{2}}$, function $K\left(\beta\right)$
is defined in \cite{Kitchatinov1993}, function $\xi_{\beta}\left(\phi\right)$
describes the longitudinal dependence of the instability, and parameter
$\beta_{cr}$ controls the instability threshold. These parameters
will be described below. As follows from the paper cited above, for $\beta\ll1$, $K\left(\beta\right)\sim1$, and
for $\beta>1$, $K\left(\beta\right)\sim1/\beta^{3}$. In this formulation,
the preferable latitude of the ``active region emergence'' is determined
by the maximum energy of the toroidal magnetic field, see Eq.(\ref{eq:buoy}).
Parameter $\beta_{cr}=0.5$ is adopted to prevent the emergence of active
regions at high latitudes. For further details concerning the model
see \cite{Pipin2018e}. Also, the python code for the model can be
found in \cite{Pipin2019}. 
Fig.~1 shows snapshots of the radial
magnetic field in the model during the maxima and minima of the dynamo
cycle. More details about the model can be found in \cite{Pipin2018d}.
 
\begin{figure}
\includegraphics[width=0.95\columnwidth]{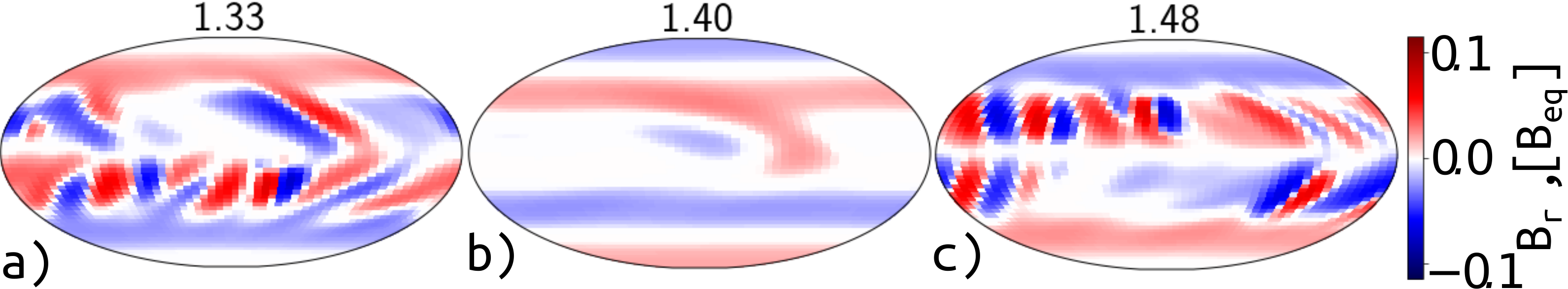} \caption{\label{fig:m1asn} {Snapshots of the radial distribution of the magnetic field
in the model for the cycle maximum (a) and (c) and for the cycle minimum (b). The time above the snapshots is given in  diffusive units.}}
\end{figure}

Note that the magnetic buoyancy is the only nonlinear effect in the model.
 Therefore, the magnetic flux loss determines the amplitude of the dynamo waves.  
The lower the parameter $\beta_{cr}$, the more efficient the magnetic flux loss and the smaller 
the magnitude of the large-scale toroidal magnetic field. The time in the model is 
measured in diffusive units. Here, we scale the dynamo period to be about 10 years.

\section{Results: QSO in sunspot data and in solar magnetograms}

In this section, we discuss the QSO manifestations in the observational
parameters. Then, we compare the observations with the results obtained
from our solar dynamo model. 
\begin{figure}
\includegraphics[width=0.99\columnwidth]{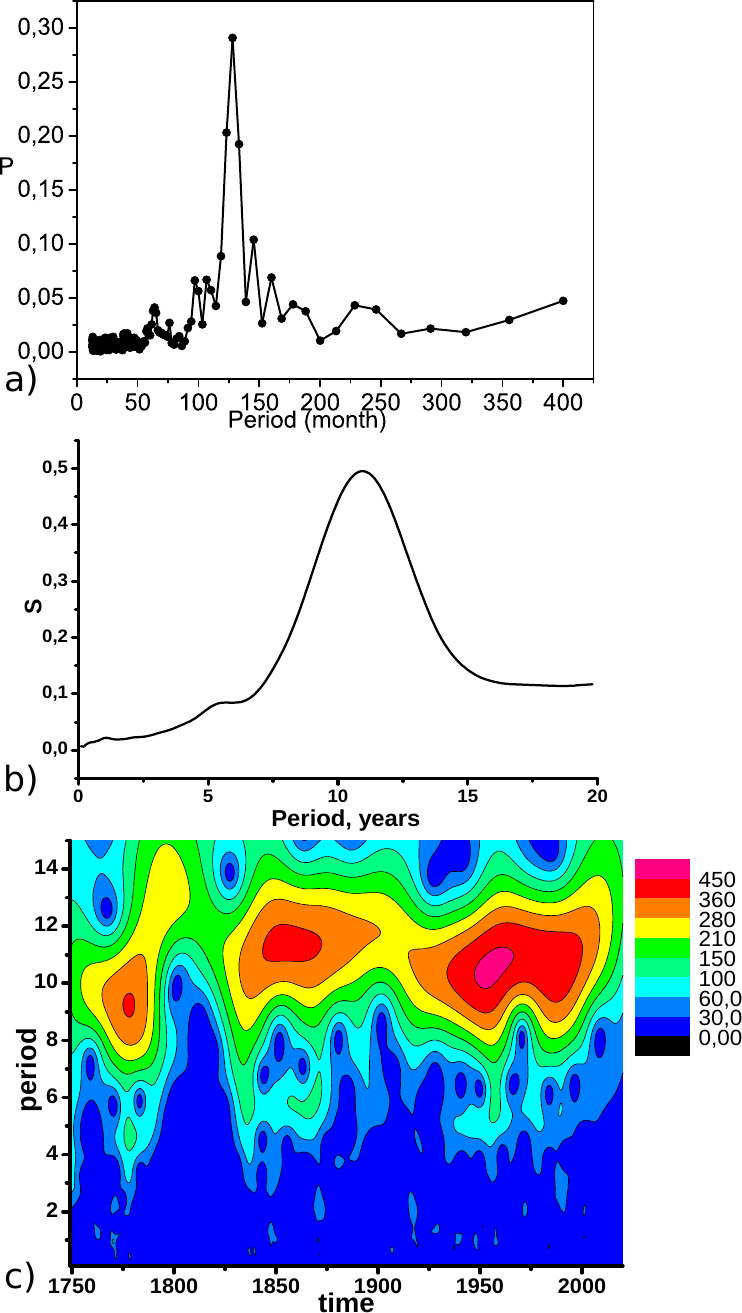} \caption{\label{Fig1}
a)Fourier spectrum for the monthly mean sunspot data
for time interval 1749 - 2019.  $P$ denotes the amplitudes of the
Fourier harmonics normalized to the sum of their moduli. The harmonic
of the maximal amplitude corresponds to the 25th harmonic, i.e., 126
months. The spectral resolution in the spectrum calculated is about
five months. Panel b) shows integral wavelet spectra for sunspot data, and panel c) shows the wavelet plane for the real part of the modulus of the sunspot wavelet coefficients.}
\end{figure}

We start with a straightforward search for QSO in the Fourier spectra
for sunspot data. Fig.~2a presents the integral Fourier spectrum for the monthly mean sunspot data 
from 1749 up to 2019. As expected, the peak corresponding to the Schwabe cycle is dominant in the plot. 
However, the point is that an additional peak near 6 years is visible, as well.

\begin{figure}
\includegraphics[width=0.99\columnwidth]{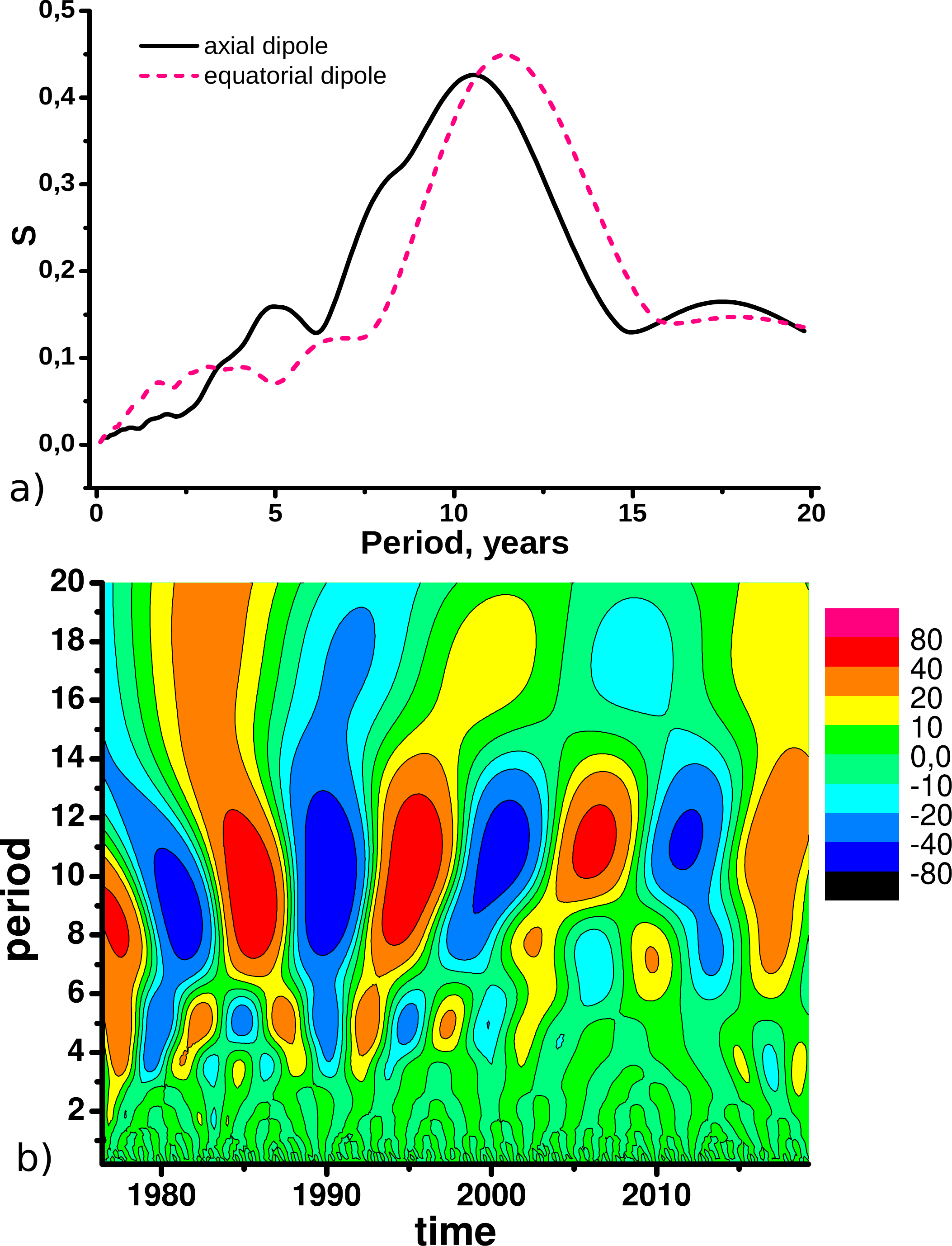} \caption{\label{Fig2}a) Integral wavelet spectra for the solar axial dipole data; b) wavelet plane for the real part of the modulus of the axial dipole
wavelet coefficients.}
\end{figure}

Fig.~2b shows the integral wavelet spectra for the sunspot time
series from 1749 up to 2019 (middle panel). Apart from the expected 11-year
peak, we recognize a much smaller peak at about 5 years. The
wavelet plane (Fig.\ref{Fig1}c) shows the long-term evolution of
the sunspot number QSO. 
One can see that QSOs have a considerable magnitude only during the epochs of the centennial maxima. 
Below, we show how this phenomenon can be explained in terms of the nonlinear dynamo model.

From the viewpoint of the solar dynamo, the sunspot data trace the
solar toroidal magnetic field.  The poloidal magnetic field is another component of the large-scale solar activity. 
The axial magnetic
dipole is a proxy parameter, which measures the strength and direction of the large-scale poloidal magnetic field.
 We employ the strength of the equatorial dipole as a proxy for the sunspot activity. 
Fig.~3 (top) shows the integral wavelet spectrum for the solar magnetic dipole. 
Since the magnetic dipole has two components of substantially different nature, i.e.,
 the axial and the equatorial dipoles, their corresponding
spectra are given separately. 
\begin{figure}
\includegraphics[width=0.99\columnwidth]{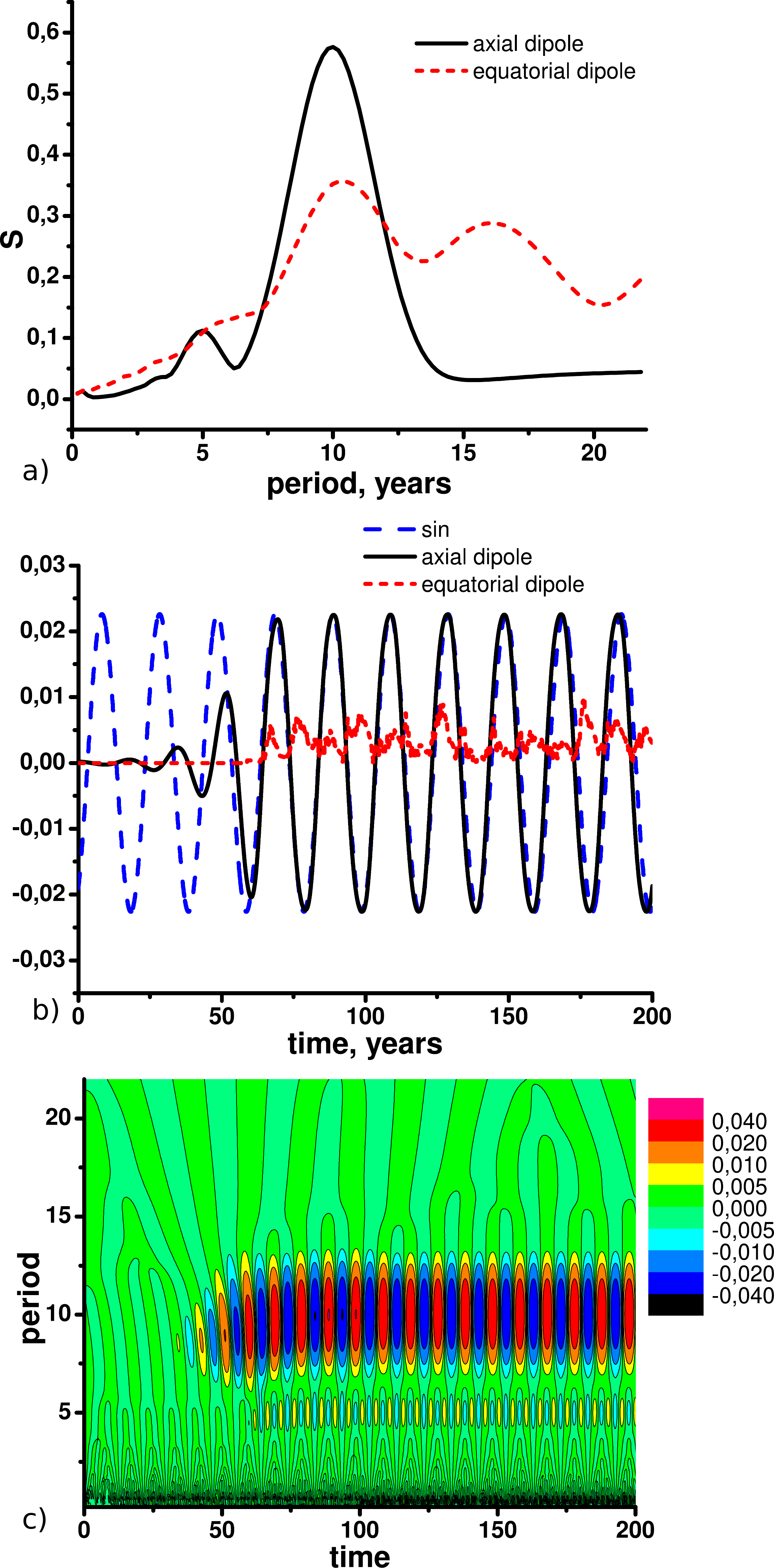} \caption{\label{Fig3} Dynamo model, upper panel: a - the integral wavelet spectra
for the axial and equatorial dipole, b - the time evolution of the axial
dipole (black line). The red line shows the absolute value of the equatorial
dipole, and the blue line shows the sinus signal; c - the wavelet plane
for the real part of the wavelet coefficients for the axial dipole.}
\end{figure}

A comparison of Fig.\ref{Fig1}b and Fig.\ref{Fig2}a shows that QSOs
with a period of 5 years are better pronounced on the plot for the axial
dipole than in the sunspot data. In contrast, the peak is absent in
the integral wavelet spectrum for the equatorial dipole (see Fig 3a).
 This empirical result sounds reasonable from the theoretical point of view,
  because it is the solar toroidal magnetic field and the axisymmetric part of the solar poloidal magnetic field 
that are involved in the solar dynamo, while the solar
equatorial dipole is something not obligatory for the solar activity cycle.

The QSO period varies from cycle to cycle both for the sunspot and
the axial dipole data sets. The period of the axial dipole QSO changed from about 5 years near the minimum of cycles 21 and 22 to 6-7 years near the minimum of cycle 23 (i.e., approximately in 2004 and 2010). Also, we see that the maxima of the axial dipole QSO are approximately located at the ascending and descending branches of the axial dipole cycle.

\section{Results: QSO in the simulations from our dynamo model}

Now, compare the results of the spectral analysis based on observational data 
with the data derived from the mean-field dynamo model. 
A preliminary analysis shows that QSO variations
in the axial dipole can be reproduced in terms of different mean-field
dynamo models, e.g., the 1D models proposed by \cite{Moss2008} and
the recent 2D model proposed by \cite{Pipin2020}. We have found out
that the presence of the non-linear dynamo saturation effect is sufficient
for the emergence of QSO both in the parameters of the toroidal magnetic
field and in the axial dipole.

For our study, we need the parameters of both the axisymmetric and
non-axisymmetric large-scale magnetic fields. For this reason, we use
the simplified version of the non-axisymmetric dynamo model. 
The model simulates the dynamo process in a thin layer deep within the convection zone. 
The effect of magnetic buoyancy seeds the bipolar active region at
a random position within the large-scale toroidal magnetic field.
This effect accounts for the escape of magnetic energy from the dynamo
region, as well.
Following \cite{Frick2020}, we use the mean density of the axisymmetric toroidal magnetic field
 as a proxy for sunspot activity:
\begin{equation}
\overline{B_{\phi}}=\frac{1}{2}\int_{-1}^{1}\left|B_{\phi}\right|d\mu,
\end{equation}
where $\mu=\cos\theta$ ($\theta$ is the polar angle), and the
total unsigned flux of the radial magnetic field is 
\begin{equation}
F_{S}=\int\left|B_{r}\right|dS,
\end{equation}
where the integration is performed over the visible hemisphere. \cite{Frick2020}
used $F_{S}$ as a tracer of the sunspot activity to study QBO.

The integral wavelet spectrum (Fig.\ref{Fig3}, upper panel) for the
absolute value of the axial dipole does contain a 5-year peak. 
The 11-year peak is present, as well. 
The equatorial dipole does not show this variation.

To clarify the nature of the QSO periodicity, we have followed
the evolution of the magnetic field starting with the tiny seed magnetic
field (Fig.~4, middle panel). It is found out that the 11-year periodicity starts at the very beginning
 of the magnetic field evolution. Quasi-sexennial oscillations gain considerable power, when the dynamo cycle becomes stationary
 as the magnetic energy reaches the nonlinear saturation state. 
This means that QSO can be considered a nonlinear effect. Note that the nonlinear saturation in the model
 is due to the magnetic buoyancy effect. The middle panel in Fig.~4
 shows the time evolution of the axial and equatorial dipoles. 
The latter looks like noise, while the time evolution of the axial dipole seems to be almost sinusoidal. 
Also, we have found out that the maximum power of the axial dipole QSO is observed at the rise
and decline of the axial dipole cycle. This dynamo-model result agrees qualitatively with
our observational findings.

\begin{figure}
\includegraphics[width=0.99\columnwidth]{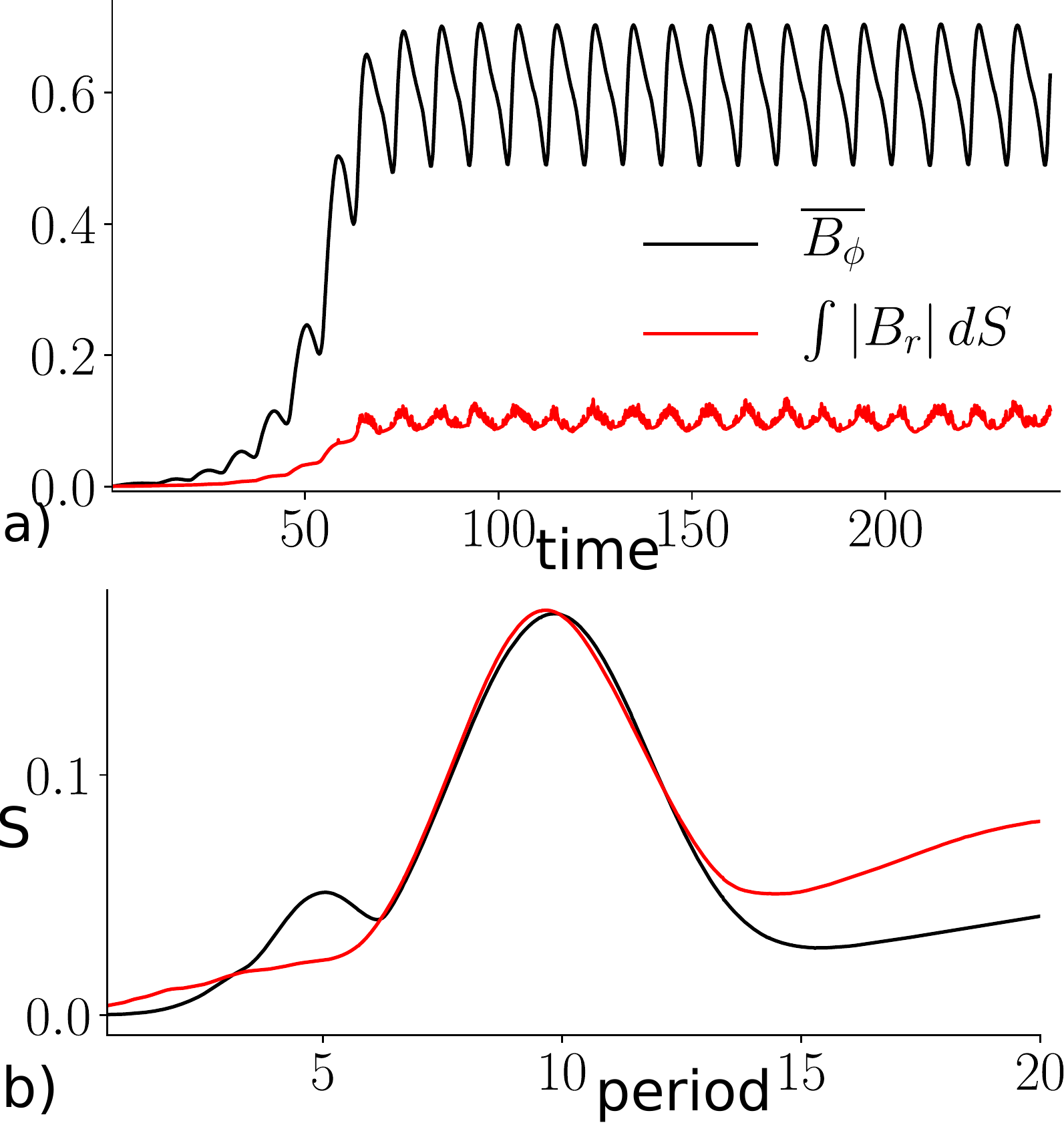} \caption{\label{Fig4}a) Evolution of the mean flux density of the toroidal
magnetic field (black line) and the total flux density of the radial
magnetic field (red line); both are shown in dimensionless units;
b) their integral wavelet spectra.}
\end{figure}

Figure \ref{Fig4} shows the evolution and wavelet spectra for the
mean flux density of the toroidal magnetic field, $\overline{B_{\phi}}$, 
the total flux density of the radial magnetic field, $F_{S}$, and
their integral wavelet spectra. The nonlinear phase of the evolution of
$\overline{B_{\phi}}$ is strictly periodic. However, the shape of
the $\overline{B_{\phi}}$ signal differs strongly from the sine signal.
Similar to the axial dipole, the integral spectrum of $\overline{B_{\phi}}$
clearly shows QSO with a period of 5 years (half a dynamo cycle).
We do not show the wavelet plane of $\overline{B_{\phi}}$. It is
qualitatively very similar to that of the axial dipole with the maximum
QSO power in the rise and decline phases of the $\overline{B_{\phi}}$
cycle.

\begin{figure}
\includegraphics[width=0.99\columnwidth]{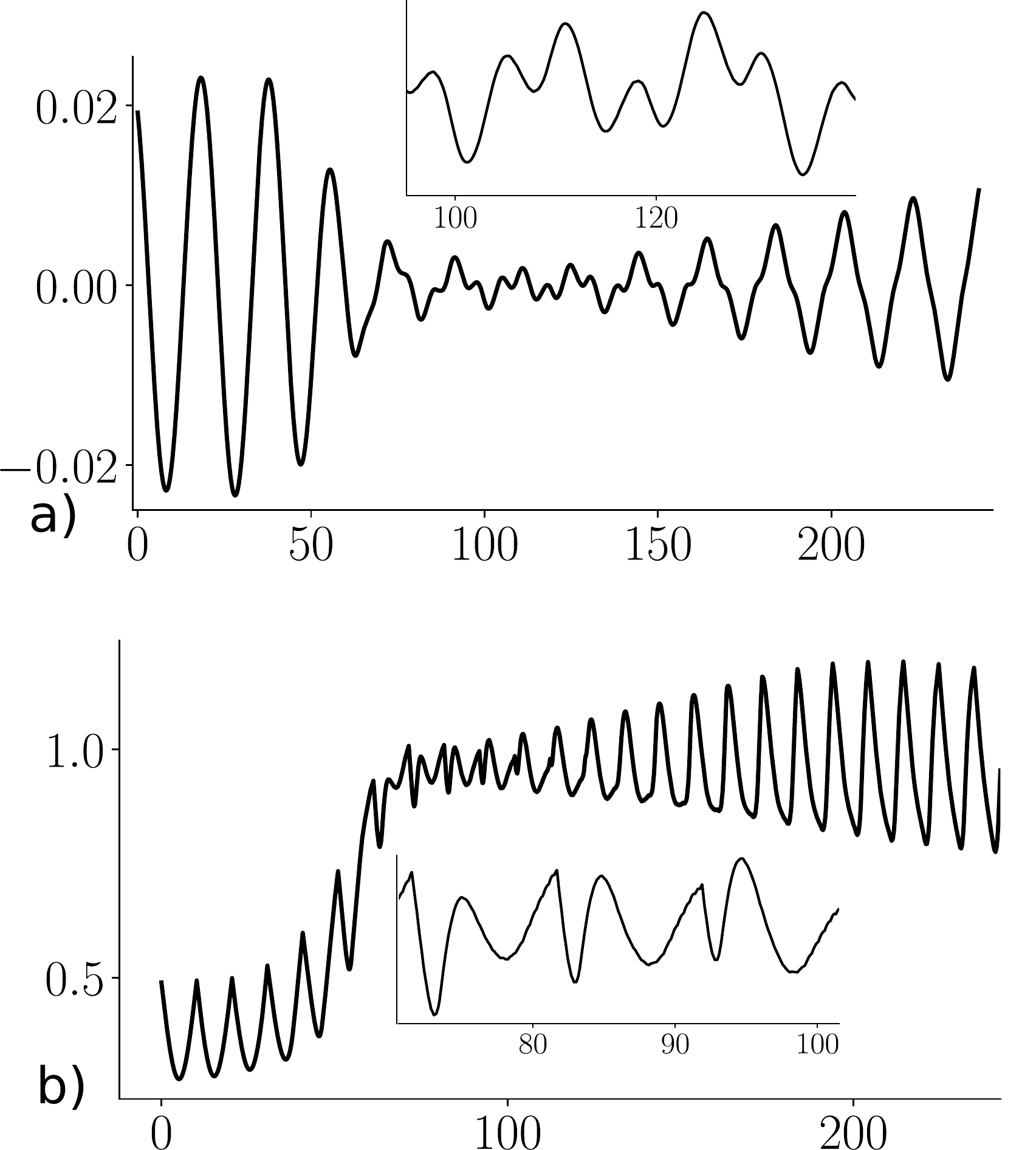} \caption{\label{Fig5}The sinus signal subtracted from the axial dipole: a)
magnetic dipole time series, b) the same for the toroidal magnetic
field. The small plots on the right present the most instructive
parts of the plots.}
\end{figure}

To separate the QSO from the main magnetic cycle, we subtract
the pure sinusoidal signal from the signal of the axial dipole. For
the toroidal magnetic field, we subtract the absolute value of the
sinusoidal signal. The results are shown in Fig.\ref{Fig5}. The resulting
signal shows tiny 5-year variations when the phase of the axial dipole
and the sine are synchronized (Fig.\ref{Fig5}, top). The oscillation
magnitude is rather small compared to the main signal. We have enlarged
the piece of the trajectory containing a clear 5-year oscillation
and have superimposed it on the main panel. The $\overline{B_{\phi}}$
parameter shows a qualitatively similar result. However, the cycle
of the toroidal magnetic field deviates strongly from a sine curve. Therefore,
the extracted QSO differs very much from the regular oscillation.

In our model, the axisymmetric toroidal field of strength $\left|{B_{\phi}}\right|>0.5$
produces bipolar regions. Evolution of the total flux of the radial
magnetic field, $F_{S}$, from these bipolar regions is shown in Fig.\ref{Fig4}
as well. We see that, in our model, only a small part of the flux
of the toroidal magnetic field is transformed into magnetic bipolar
regions. The integral wavelet of $F_{s}$ does not show QSO (Fig.\ref{Fig4}b).
This does not mean that QSO in $F_{S}$ are absent at all.

\begin{figure}
\includegraphics[width=0.99\columnwidth]{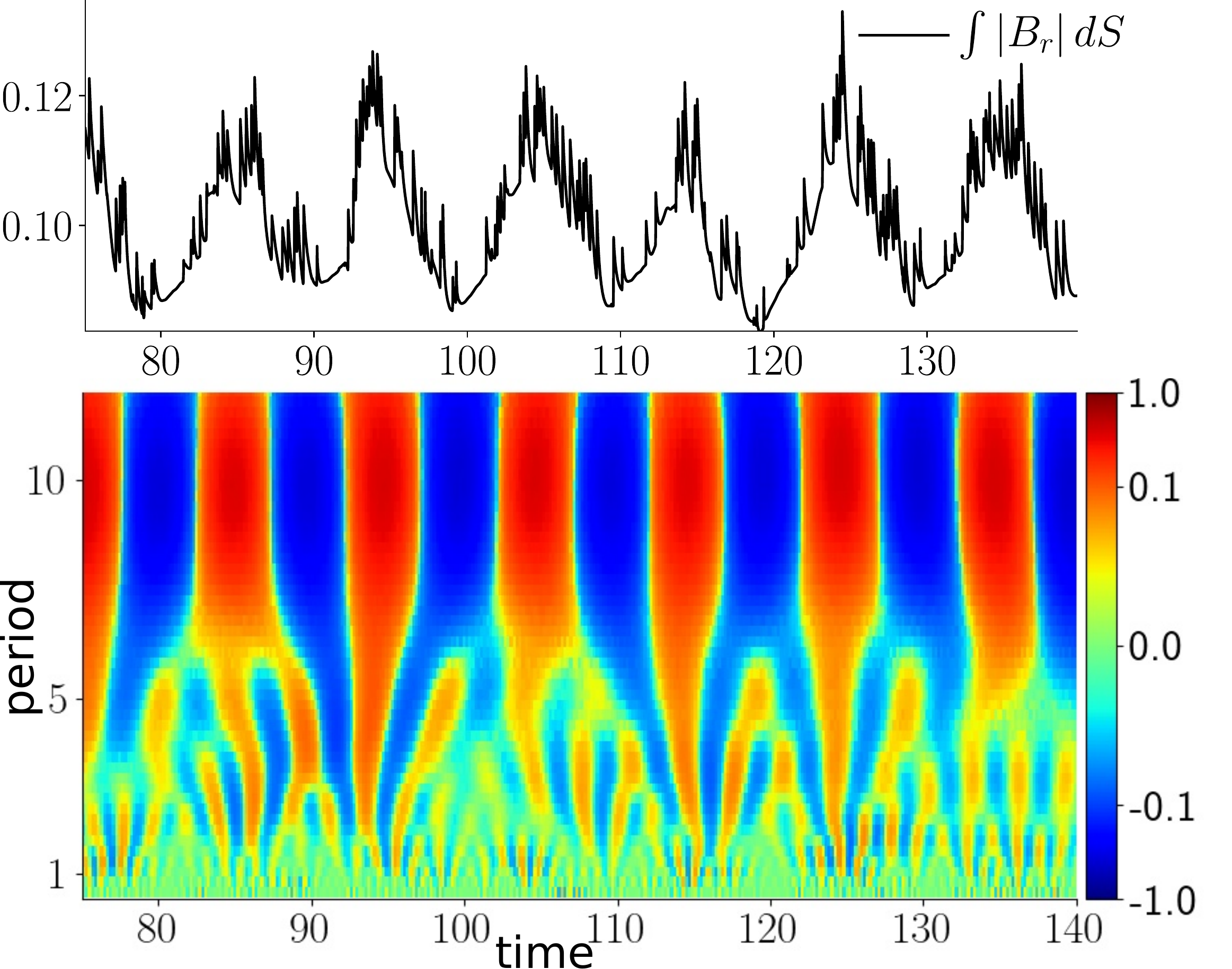} \caption{\label{sflux}a) Evolution of the total flux of the radial magnetic
field, $F_{S}$; b) wavelet plane for the real part of wavelet coefficients
of $F_{S}$.}
\end{figure}

Figure \ref{sflux} shows a piece of the $F_{S}$ trajectory
for the nonlinear stage of the dynamo evolution and its wavelet plane
for this time interval. We see, that the shape of the $F_{S}$ cycle
differs from the cycle of $\overline{B_{\phi}}$. The $F_{S}$ spectrum
contains variations in the range of periods of both QBO and QSO. However,
these intervals seem to have no gap in between. Therefore the QSO
power is not pronounced in the integral spectrum. It is interesting
to note that, in the given time interval, the greatest power of QSO with a 5-year period was 
observed near the minima of the $F_{S}$
trajectory. This differs from the behavior of $\overline{B_{\phi}}$.

\section{Discussion and Conclusions}

\begin{figure}
\includegraphics[width=0.9\columnwidth]{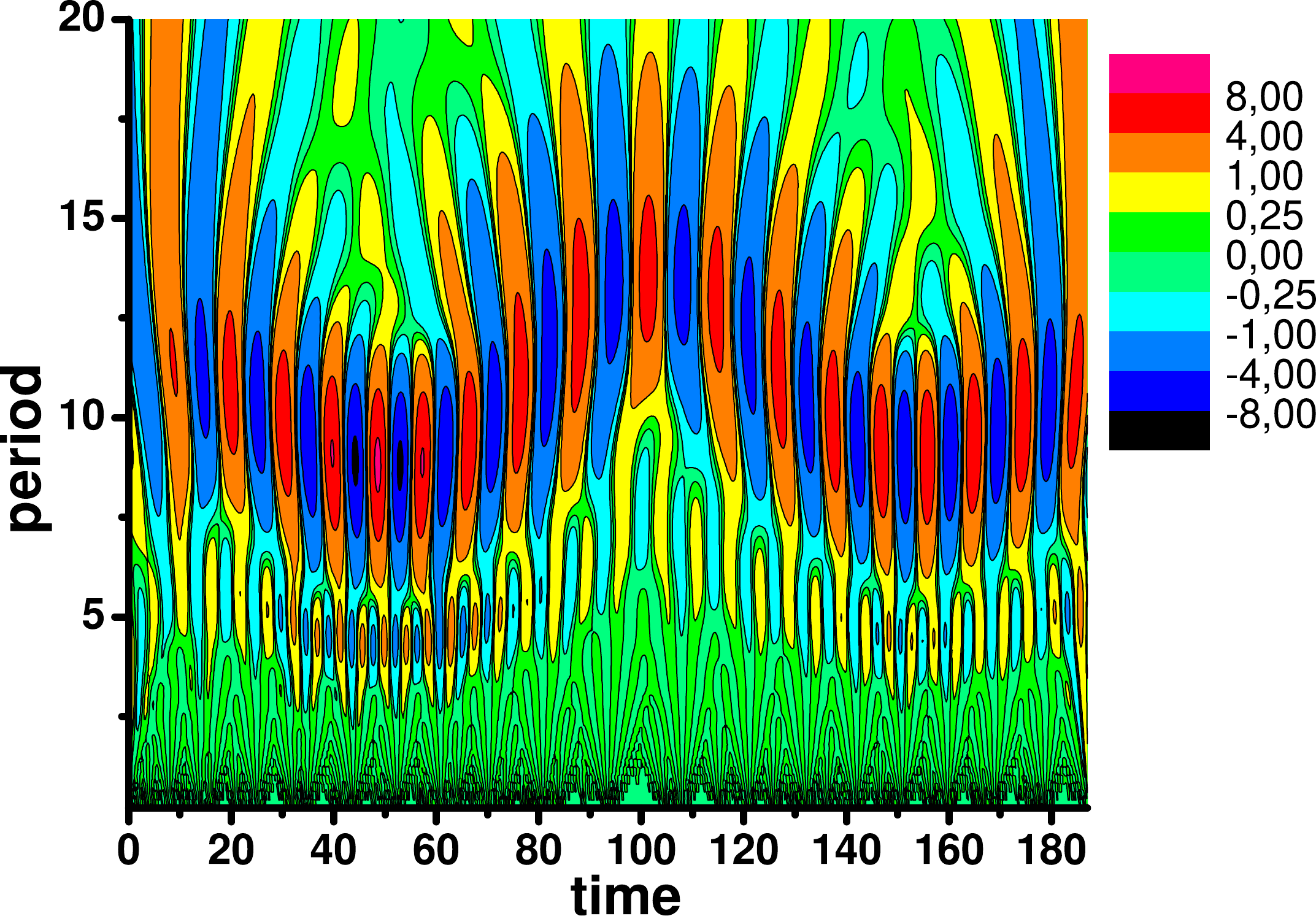} \caption{\label{C4}2D wavelet spectra for the model with a variable $\alpha$
effect.}
\end{figure}

We have isolated QSO hidden in the phenomenon of solar activity using sunspot data and 
the record of solar magnetic dipole.
These time series trace the magnetic field components involved in
the solar dynamo, which maintains the main solar cycle, i.e., the
toroidal and poloidal components.
We have reproduced a very similar behavior using a relatively simple solar dynamo model, 
which is, however, sophisticated enough to include the non-axisymmetric solar magnetic field.
 We have uncovered that in our model, the axial dipole and the mean density of the
toroidal magnetic field display QSO only during the nonlinear stage of
the dynamo evolution. As directly follows from the analysis, the axial-dipole
QSO can be interpreted as variations in the shape of the axial dipole
cycle due to the non-linearity of the underlying dynamo process. In our
model, the main non-linearity comes from the magnetic buoyancy effect.
This effect results in a strongly asymmetric shape of the toroidal
magnetic field cycle. In the model, the poloidal magnetic field is
produced from the toroidal field by the $\alpha$ effect. The shape
of the dipole cycle becomes asymmetric because of the asymmetric shape
of the $\overline{B_{\phi}}$ cycle. Therefore, the axial-dipole QSO
is inherent to the QSO of $\overline{B_{\phi}}$. The toroidal magnetic field in our model, besides
participating in the dynamo process,  produces
bipolar regions. These bipolar regions modulate the total flux of the radial magnetic field on 
the surface, $F_{S}$. This parameter
does not show QSO in the integral wavelet. This may be due to nonlinear
multi-scale $B^{2}$ processes in the evolution of the non-axisymmetric magnetic fields that are 
initiated by the formation of bipolar regions.
It is noteworthy that the QSO in sunspots are less pronounced than
the axial-dipole QSO. This empirical result may be related to the QSO saturation effect
similar to that found for the $F_{S}$ parameter.

Another possible explanation is that the sunspot data series 
are much longer than those for the axial dipole. Note that observations show
long-term variation in the QSO parameters. A weak axial-dipole cycle
tends to have a longer QSO period. For example, the period of the
axial-dipole QSO was shorter in cycles 21 and 22 than in cycles 23
and 24 (see, Fig.\ref{Fig2}). This conclusion is not robust because
we have actual observational data for only a few cycles. We find that the given
conclusion can be drawn using the time series from dynamo simulations
that show a long-term variability. We checked it using the time series
in the recent model by \cite{Pipin2020} (i.e., model C4). 
Note that in this model, the long-term variation of the magnetic activity was derived from 
 \textit{regular} centennial variations of the
$\alpha$ effect (for more details see the paper cited above). 
Variations in the period of the axial dipole cycle and QSO variations 
in that dynamo model outputs are shown in Figure \ref{C4}.
 We see that the shift of the
cycle period results in the corresponding shift of the QSO period.
This is in qualitative agreement with the results shown in Fig. 2c earlier.
The model shows the period of the dipole cycle to increase to about
15 years during the grand minimum. The corresponding QSO almost disappear during
this period. The model shows a similar behavior of the long-term
evolution of QSO of the mean flux density of the toroidal magnetic
field. It is noteworthy that the integral of the wavelet spectrum
shown in Fig.\ref{Fig2} shows no considerable power in the QSO range
because of dispersion. Bearing in mind our mechanism of QSO, we suggest
that the analysis of longer observational time-series of the axial
dipole may reveal saturation of the QSO power in the case of strong
variations in the axial dipole cycle.

We admit that other origins for a given QSO can exist as well. 
For example, in the flux-transport models the evolution of the axial
dipole parameters can depend on variations in the meridional circulation.
This mechanism should be studied separately. Note, that the model
by \cite{Pipin2020} is the state of the art non-kinematic mean-field
axisymmetric dynamo model. The results concerning the QSO derived
from this model need further investigation. We mention them here to
support our primary conclusion that the non-linear dynamo effects are the
main source of QSO. A QSO has half the period (twice the
frequency) of the basic cycle; so the form of nonlinearity must
ensure this. The magnetic buoyancy is the  $\left|B\right|^{2}$
effect. Also, we have analyzed how different kinds of the $\alpha$-effect
 nonlinearities affect the QSO. For example, we have studied
dynamo models with the so-called "algebraic"
and dynamic $\alpha$ - effect nonlinearity. 
The latter results from the magnetic helicity conservation. 
We find that these nonlinearities
result in anharmonic shapes of the toroidal magnetic field cycles. 
Similarly to the magnetic buoyancy, these nonlinear dynamo effects have 
the  $\left|B\right|^{2}$ - type dependence on the strength of the large-scale magnetic field. 
Therefore, they can result in modulating the $\alpha$ effect, as well as in the escape of magnetic field 
 with the doubling of frequency relative to the original frequency of the mean magnetic field strength. 
This causes anharmonic shapes of the toroidal magnetic field cycles (see, \cite{Baliunas2006}), 
and it makes the QSO features
robust for the stationary phase of the dynamo evolution in our models.

 We note, however, that some other parameters are not robust. For example, the 11-year cycle
in the equatorial dipole is not well seen in the models with the "algebraic", $B^{-2}$,
 non-linearity of the $\alpha$- effect. The above discussion
does not exclude that some QSO power can stem from random fluctuation
in the shape of the activity cycles. To distinguish between the random
and regular mechanisms generating the QSO, we need to extend our study.
This can be done using more sophisticated data mining tools and dynamo
models with random fluctuations.

A related point is that nonlinear dynamos are often invoked
to explain lower frequencies in the records of solar activity (i.e.,
modulation of the basic cycle on longer timescales). This approach
is a natural consequence of nonlinear interactions between the
basic parity modes of the dynamo-generated magnetic field (see 
\citealp{Ivanova1976,Brandenburg1989,Sokoloff1994,Knobloch1998,Weiss2016}).
It was found that if two different parity modes (e.g., symmetric
and antisymmetric about the equator) have both close dynamo periods
and dynamo excitation thresholds, then their nonlinear interactions can
result in the modulation of the total magnetic energy with a long
period determined by the difference between the mode frequencies.
Also, the nonlinear interaction of a large-scale magnetic field
and large-scale flows can result in modulation of the magnetic activity 
(see, \citealp{Brandenburg1991,Knobloch1998,Pipin1999,Kuker1999}). In
both cases, the time scale of the parity mode interactions and the relaxation
of magnetic perturbations affect the large-scale flows 
over time intervals longer than the main dynamo period. This is
opposite to the effects of the magnetic cycle shape variations discussed
in this paper. Certainly, it is a very attractive and challenging task
to study the nonlinear dynamo mechanisms on  short and long time
scales as  part of the continuous transport of the magnetic and kinetic
energy in the system. It looks plausible that these phenomena should
be presented in various nonlinear systems, and their further investigation
seems to be promising.

To sum up, we can draw the following conclusions. Observations show
the existence of QSO both in the sunspot number
 record and in the proxies of the global magnetic field of the
Sun, e.g., the evolution of the solar axial dipole. 
The dynamo models show
that these QSO result from the nonlinear, anharmonic shape of the dynamo cycle. 
The inter-scale dynamics initiated by the sunspot formation at the solar surface 
and the long-term dynamo variations result in long-term variations of QSO. 
This work opens a new window for future study.

\textbf{Acknowledgements}

The authors are grateful to the WSO teams for a free access to their
data. The work was supported by RFBR grants No. 20-02-00150, 18-02-00085
and 19-52-53045. VVP did the dynamo simulations as part of the scientific
project FR II.16 of ISTP SB RAS.

We are also grateful to Dr. W.Sun (Harvard-Smithsonian Center for Astrophysics) for critical reading of the manuscript.

 Data Availability Statements.

The observational data underlying this article are available at http://wso.stanford.edu/
and http://sidc.oma.be/silso/datafiles

The python code of the dynamo model and it results are available at zenodo.org
\cite{Pipin2019}.

\bibliographystyle{mnras}
\input{5year-n.bbl}
 
\end{document}